\documentclass[aps,prb,twocolumn,groupedaddress]{revtex4}

\usepackage{graphicx}
\usepackage{bm}

\begin{document}


\title{Static and dynamical dipolar / strain fluctuations
in perovskite ferroelectric relaxors}


\author{Y. Yamada and T. Takakura}
\affiliation{Advanced Research Center for Science and Engineering,
Waseda University, 3-4-1 Okubo, Shinjuku-ku, Tokyo 169-0072 Japan}


\date{\today}

\begin{abstract}
We develop a theory to study the characteristics of dipolar / strain 
fluctuations in perovskite relaxors. In addition to the soft TO and TA phonons, we take into account another freedom of motion associated with the random 
hopping of $\mathrm{Pb^{2+}}$ ions between the off-center sites around the high symmetry 
corner site, thus constructing a coupled TO-TA-pseudospin model to describe 
the perovskite relaxors. It is shown that there is a possibility that prior 
to the on-set of instability of the uniform TO-mode (soft-mode), instability 
of TA mode with nanometer scale modulation $(\mathbf{q}\sim 0.1\mathbf{a}^{*})$ takes place, which produces static heterogeneous structure concerning 
polarization as well as shear strain. This seems to suggest the intrinsic 
origin of the heterogeneity in relaxors visualized as random distribution 
of PNR. The phonon spectral density distribution has also been investigated. 
It is shown that when the relaxation time of random hopping of 
$\mathrm{Pb^{2+}}$ ions is comparable to the TO phonon frequency at 
$\mathbf{q}\sim 0.1\mathbf{a}^{*}$, the calculated phonon spectral density 
reproduces the characteristic features called `waterfall'. 
\end{abstract}


\maketitle

\section{Introduction}

Recently, the unique properties of ferroelectric relaxors have become 
one of the central topics in solid state physics. The diffuseness of 
the dielectric response against temperature variation has been considered 
to be due to the randomness of the system created by the random occupation of 
B-site ions with different valences. In fact, the basic physics of relaxors 
has been mainly discussed in terms of `random field' to stabilize a glassy 
state analogous to spin glasses.

One of the intriguing problems to be answered is the role of 
$\mathrm{Pb^{2+}}$ -ions on A-site. In spite of the efforts to eliminate 
Pb ion based on technological reasons, hitherto known perovskite relaxors 
seem to be restricted to Pb compounds, which suggests that at least part 
of the unique properties would be related to Pb ions at A-site.

From experimental view point, there exist a few unique features exhibited 
by relaxors both in structural as well as lattice dynamical aspects 
providing some `key concepts' to describe characteristic features of relaxors. 
The key concept to discuss the static structure of relaxors is the so-called 
PNR (polar nanoregion). That is, the averaged cubic symmetry is locally broken 
in the temperature region where dielectric constant exhibits broad maximum. 
The overall structure is thus visualized as a random distribution of PNR 
embedded on cubic parent phase.

We notice that there are a few materials which show similar intrinsic 
heterogeneity in the vicinity of phase transition point. A group of 
bcc-based alloys called shape memory alloys, which undergo martensitic 
transformation, exhibit heterogeneous structure where the `embryo's or 
microdomains of martensite are embedded on the austenite (bcc) matrix 
over a wide temperature range. More recently, a group of perovskite 
manganites called CMR substances, which undergo metal-insulator transition, 
have been noticed to develop heterogeneous structure\cite{Shimomura} where 
microdomains of metallic phase are coexisting with insulator phase. 
The close relationship between CMR substances and relaxors was already 
pointed out by Kimura et al.\cite{Kimura}  

Onuki\cite{Onuki} investigated the origin of the stability of two-phase 
coexistence in alloys and pointed out that the coupling between the order 
parameter to the local strain is essential to stabilize the heterogeneous 
structure. Later, Yamada and Takakura\cite{Yamada1} also arrived at the same 
conclusion in the case of CMR substances.

In this connection, the recent neutron scattering study on PMN 
by Hirota et al.\cite{Hirota} seems to be very suggestive. They carried out 
the dynamical structure analysis of the diffuse scattering and concluded that 
the displacement pattern of each ion in the unit cell contains considerable 
amount of CM (center of mass) non-conserving components. In the language of 
phonon modes, it means that the normal coordinate of the condensing mode is 
given by a linear combination of TO-mode and TA-mode suggesting 
the possibility of strong polarization-strain coupling.

On the other hand, the key concept to characterize the dynamical aspect of 
relaxors is so-called `waterfall' in phonon spectrum. 
Gehring et al.\cite{Gehring1} carried out the pioneering neutron 
scattering study on PZN-8PT and found out that the observed high intensity 
ridge of the neutron spectrum did not follow the expected TO phonon 
dispersion at $q\leq 0.2{\bf a}^{*}$. Instead, it falls down vertically 
to precipitate onto TA dispersion. Similar features are successively 
observed in various relaxors including PMN and 
PZN,\cite{Gehring2,Shirane1,Tomeno} indicating that waterfall is indeed 
the unique lattice dynamical characteristic of relaxors.  

Gehring et al.\cite{Gehring3} analyzed the neutron spectrum of PZN based 
on the `mode coupling' treatment which was utilized 
by Harada et al.\cite{Harada} to explain the phonon spectrum of 
$\mathrm{BaTiO_{3}}$. They concluded that in order to reproduce the observed 
spectrum, the TO phonon width should show an abrupt change at $q\sim0.1a^{*}$ 
where waterfall takes place. They claim that the abrupt change is caused 
by the random distribution of PNR in the medium, thereby the propagation of 
the lattice wave with wave length longer than the average size of PNR is 
impeded. While this viewpoint is very attractive, whether the static 
heterogeneity such as PNR will cause waterfall type anomaly or not is 
unclear since a static entity will only give rise to large momentum transfer 
of the phonon without causing any energy transfer from phonons to other 
freedom of motion and eventually to heat bath. In order to establish effective 
channels of energy transfer, some dynamical entity to which phonons are 
coupled would be needed.

We consider that the configurational freedom of motion of 
$\mathrm{Pb^{2+}}$-ion will provide such possibility. It is 
known\cite{Fujishiro} that in relaxors the instantaneous equilibrium 
position of $\mathrm{Pb^{2+}}$ ion is slightly shifted from the high symmetry 
corner site and is making random hopping motion between the equivalent 
off-center sites to recover cubic symmetry on average. Formally, such freedom 
of motion can be expressed by a stochastic pseudospin variable. Based on these 
considerations we construct a suitable model of the relaxor which is 
characterized by `coupled TO-TA-pseudospin system'. In the next section, we 
discuss the TO-TA coupling within the framework of `quasi-harmonic' 
approximation which was utilized by Axe, Harada and Shirane (AHS)\cite{Axe} 
in order to analyze the anomalous TA dispersion in $\mathrm{KTaO_{3}}$. When 
applied to PMN, this treatment suggests the possibility to stabilize a 
heterogeneous static structure. In section 3, we discuss the neutron 
scattering spectra including waterfall anomaly by taking into account of 
random hopping of $\mathrm{Pb^{2+}}$  ions. The last section is devoted to 
conclusions and discussions. 

\section{Static polarization / strain fluctuations 
-- origin of polar nanoregion --}

According to the analysis of the diffuse scattering intensities observed 
in PMN around various reciprocal lattice points by Hirota et al.,\cite{Hirota} 
the structure factor of the condensing mode includes substantial amount of 
atomic displacements which does not conserve the center of mass (CM) of 
the unit cell. Since the optical modes at $q=0$ should satisfy the condition 
of CM-conservation of the coordinates, the above observation seems to suggest 
that in PMN, there would be strong coupling between the soft optical mode 
and the transverse acoustic mode around $q=0$, so that the normal coordinates 
include considerable amount of CM-nonconserving displacements. In this 
connection, we notice that there are a few perovskite ferroelectric materials, 
which exhibit anomalous behavior in the TA branch upon softening of the TO 
branch. Fig. 1 shows typical behavior in the case of 
$\mathrm{KTaO_{3}}$.\cite{Comes} It is seen that as the TO mode softens, 
the dispersion of the TA branch exhibits anomalous `concave' curve in a limited $q$-region of $0.02 \leq q \leq 0.2$. It is noticeable that the dispersion 
shows no `softening' in the vicinity of the reciprocal lattice point 
within $q \leq 0.02$. Such anomaly in the TA dispersion also indicates 
the possibility of strong coupling between TO- and TA- modes in perovskite 
ferroelectric systems.

                                                                    
\begin{figure}[h]
\begin{center}
\includegraphics[width=3.7cm,height=9.5cm]{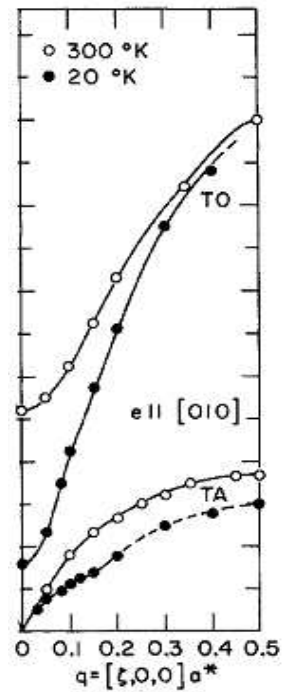}
\end{center}
\caption{The TO and TA phonon dispersions of $\mathrm{KTaO_{3}}$ at 300K and 20K given in ref. 14. As the TO branch `softens' upon lowering 
temperature, the TA branch exhibits anomalous concave curve in a limited 
$q$-region of $0.02 \leq q \leq 0.25$. Notice the dispersion shows no anomaly 
in the direct vicinity of the zone center $(q=0)$ within $q \leq 0.02$.}
\label{fig:1}
\end{figure}

In 1970, Axe et al.\cite{Axe} discussed the anomalous behavior of TA branch in  $\mathrm{KTaO_{3}}$ in the framework of `quasi-harmonic coupling' treatment. 
They express the dynamical matrix $\mathbf{D(q)}$ of the harmonic potential 
by taking the normal coordinates at $q=0$ as the basis functions. 
The off-diagonal matrix elements $D_{ij}(q)$ may then be considered as the 
mode-mode coupling energy between the $i'$th and $j'$th modes. By taking only 
TA-TO coupling into account the phonon properties (characteristic frequency and the corresponding normal coordinates) are explicitly given by solving the 
following secular equation:

\begin{eqnarray}
  \left|
    \begin{array}{cc}
	  \omega^{2}_{0}(T)+F_{11}({\mathbf q})-\omega^{2}({\mathbf q}) & 
	  F_{12}({\mathbf q}) \\
	  F_{21}({\mathbf q}) & F_{22}({\mathbf q})-\omega^{2}({\mathbf q})
    \end{array}
  \right|
    =0, \nonumber \\
    i
   \left\{
	 \begin{array}{l}
		= 1 :\mathrm{TO} \\
		= 2 :\mathrm{TA}
	 \end{array}
   \right.
\end{eqnarray}


\noindent where $\omega_{0}(T)$ is the soft mode frequency at $q=0$ which is 
the only temperature dependent quantity. $F_{ij}(\mathbf q)$ are to be 
expanded in terms of $\mid \mathbf{q} \mid $ as:

\begin{eqnarray}
  F_{ij}({\bf q})=f_{ij}^{(2)}({\bm \kappa})q^{2}
  +f_{ij}^{(4)} ({\bm \kappa})q^{4},        
\end{eqnarray}

\begin{eqnarray}
  {\bm \kappa}={\bf q} /\mid {\bf q} \mid,
\end{eqnarray}

\noindent along a specific direction ${\bm \kappa}$. The diagonal 
elements $f_{ij}^{(v)}({\bm \kappa})$'s may be obtained by comparing 
with the experimental dispersion curves at high temperatures where no 
anomaly is observed. In particular, $f_{22}^{(2)}({\bm \kappa})$ is directly 
given by the elastic constant for shear strain along 
${\bm \kappa}$-direction. The only essential parameters to be fixed 
are $f_{ij}^{(v)}$'s with $i\neq j$ describing the coupling between TO- and 
TA- modes. Axe et al.\cite{Axe} analyzed the experimental results 
on $\mathrm{KTaO_{3}}$ and successfully reproduced the TA, as well as TO, 
dispersions at various temperatures by fitting these two parameters.

We further notice that the off-diagonal matrix elements of the transformation 
matrix of the basis functions, $\mathbf S(\mathbf q)$, become comparable to the diagonal ones in the $q$-region where the TA branch exhibits anomalous 
behavior. For instance, $\mathbf S(\mathbf q)$ at $\mathbf {q} =(0,1,0,0)$ in 
$\mathrm{KTaO_{3}}$ is given by,

\begin{eqnarray}
  \mathbf{S}(0.1,0,0)=
  \left(  \begin{array}{cc}
  0.880 & 0.475 \\ -0.475 & 0.880
  \end{array} \right).
\end{eqnarray}
                   
\noindent This means that in the $q$-region where anomaly in TA dispersion 
takes place the normal coordinate of TO-mode contains large amount of uniform 
translational displacements, and vice versa. (TA-mode contains large amount 
of pure optical displacements.) This behavior reminds us the characteristics 
of the dynamical structure factors as pointed out by Hirota et al. in the 
case of PMN.

It is noticeable that $\mathrm{PbTiO_{3}}$, the prototype material of 
perovskite relaxors, was reported\cite{Shirane2} to exhibit the same type 
anomaly in TA dispersion. Therefore, it would not be unreasonable to assume 
that perovskite relaxors also belong to the materials which experience strong 
TO-TA coupling.  Unfortunately, such anomaly can not be proved directly by the 
observation of neutron spectra in relaxors because of the extraordinarily large damping characterized by the `waterfall' phenomena. 

At this stage, we try to construct the `hypothetical' dispersion curves of 
PMN within the `quasi-harmonic' coupling formalism by completely neglecting 
the anharmonicity of the potential to cause damping. Among the parameters 
defined in eq. (2), $f_{22}^{(2)}({\bm \kappa})$ is unambiguously given 
by the observed elastic constants $c_{44}$ and $1/2(c_{11}-c_{12})$ for 
${\bm \kappa}$ // [100] and  ${\bm \kappa}$ // [110] 
respectively.\cite{Yushir} The remaining $f_{ij}^{(v)}({\bm \kappa})'$s are 
assumed to be isotropic (independent of ${\bm \kappa}$) and to take the 
same values as those for $\mathrm{KTaO_{3}}$ given by Axe et al.. 
      
The results of calculation are given in Fig. 2. The dispersion of TA [100] 
shows the expected concave curve. In contrast, the dispersion of TA [110] 
exhibits a `dip' around $q \sim 0.1$. Moreover, by subtle adjustment of the 
parameters, the dip becomes more pronounced so that $\omega _{TA} \to 0$ at 
$q=0.12$. That is, the TA mode has become `condensed' to form a modulated 
static structure with modulation period 
$\lambda _{0}=2\pi/q_{0}=\frac{a}{0.12}$.

\begin{figure}[h]
\begin{center}
\includegraphics[width=8.0cm,height=6.0cm]{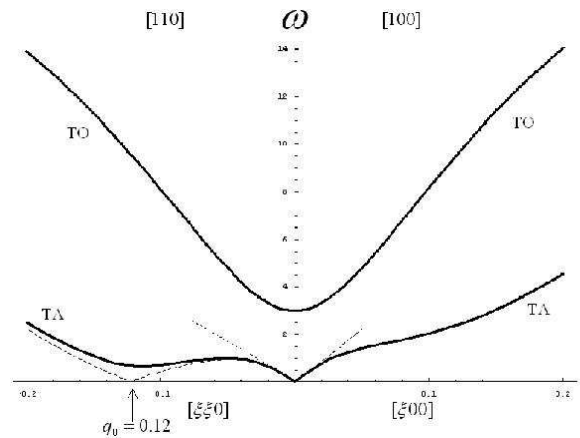}
\end{center}
\caption{The calculated `hypothetical' phonon dispersions in 
PMN. The dashed lines are the asymptotic behavior at $q \simeq 0$ determined 
by the observed elastic constants\cite{Shirane2} along [100] and 
[110]-directions. The dotted curve corresponds to the case when the 
coupling parameters are slightly modified from the values determined for 
$\mathrm{KTaO_{3}}$.}
\label{fig:2}
\end{figure}

The transformation matrix at $q_{0}=0.12$ is calculated to give 

\begin{eqnarray}
S(0.12,0.12,0)=
  \left(  \begin{array}{cc}
  0.895 & 0.445 \\ -0.445 & 0.895
  \end{array} \right).
\end{eqnarray}

\noindent Therefore the local structure of the condensing TA mode is expressed 
as the linear combination of uniform translational and pure optical 
displacements with the ratio of $S_{21}/S_{22}=0.50$.

Physically, this implies that the TA [110] mode with the wave length of 
nanometer order, $\lambda _{0}$, has the tendency to become unstable prior 
to the condensation of the uniform $(q=0)$ TO-mode due to the quasiharmonic 
coupling. Hence the system eventually stabilizes a polarization / strain 
modulated structure with wave length $\lambda _{0}$. As has been pointed out, 
such characteristics of the phonon dispersions are not able to be observed 
by inelastic neutron scattering experiment obscured by the large damping. 
On the other hand, X- ray diffuse scattering intensity, which is proportional 
to the instantaneous correlation of fluctuations, seems to give some 
indirect information on the characteristic frequency $\omega({\bf q})$.

The X-ray intensity due to the excitation of the phonons belonging to 
$\lambda$'th branch is given, irrespective of the property of 
damping,\cite{dissipation} as

\begin{eqnarray}
  I^{\lambda}({\mathbf K})=
  \frac{1}{\omega_{\lambda}^{2}({\bf q})}
  \mid F_{\lambda}({\bf K})\mid ^{2},
\end{eqnarray}

\noindent where $F^{\lambda}({\bf K})$ is the dynamical structure factor of 
the $\lambda$'th mode. In the present system, where TA-mode, rather than TO, 
is assumed to become extremely soft, we may express the X-ray diffuse intensity around each Bragg point, ${\bf K}_{h}$, in the form: 

\begin{eqnarray}
  I({\bf K})\cong \frac {1}{\omega_{TA}^{2}({\bf q})}
  \mid F_{TA}({\bf K})\mid,
\end{eqnarray}

\begin{eqnarray}
  F_{TA}({\bf K})\cong S_{22}F_{TA}^{0}({\bf K}_{h})
  +S_{21}F_{TO}^{0}({\bf K}_{h}),
\end{eqnarray}

\noindent where $F_{TA}^{0}({\bf K}_{h})$ and $F_{TO}^{0}({\bf K}_{h})$ are the dynamical structure factors for the pure translational, and pure optical 
displacements at $q=0$ respectively. Fig. 3(a) shows the calculated X-ray 
diffuse scattering intensity distributions around three Bragg positions in 
comparison with the observed X-ray results by You et al..
(Fig. 3(b)).\cite{You} While overall characteristics of anisotropic 
distribution are reproduced qualitatively, a remarkable discrepancy is seen 
in the diffuse pattern around (400). The observed distribution does not show 
any clear existence of `satellites' which are shown in the calculated 
contour. This point will be discussed later.

Although it is not shown explicitly in the figure, the relative intensities 
of the diffuse pattern reflects the degree of mixing of the modes, through 
the structure factors. Hirota et al.\cite{Hirota} gave the relative 
displacement for each ions in the unit cell which are divided into 
CM-conserving (TO-like) and CM-nonconserving (TA-like) components 
(See eq. (6) in ref. 5). Using the table, the experimental value 
of $S_{21}/S_{22}$ is obtained as 0.63, which should be compared with the 
calculated value of 0.50 (See eq. (5)).

\begin{figure}[h]
\begin{center}
\includegraphics[width=8.0cm,height=12.0cm]{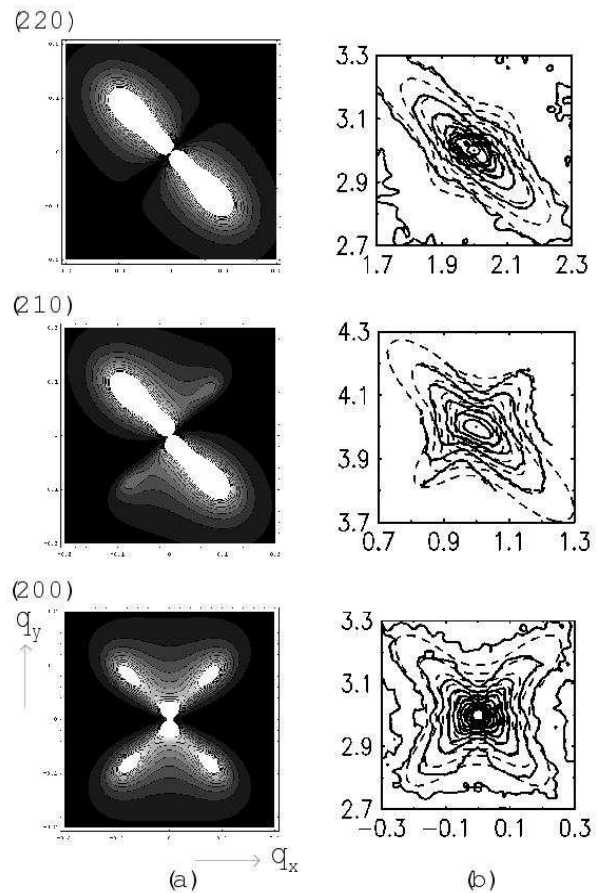}
\end{center}
\caption{(a) Calculated X-ray diffuse intensity distributions 
around various types of Bragg reflections, which are compared with the observed intensity distributions by You et al.\cite{You} reproduced in (b).} 
\label{fig:3}
\end{figure}

\section{Dynamical polarization / strain fluctuations --origin of waterfall--}

The neutron spectrum of PMN along [100] in the temperature region where 
`waterfall' phenomenon takes place has been analyzed by Gehling 
et al.\cite{Gehring3} based on the coupled mode treatment utilized by 
Harada et al. to discuss the asymmetric spectra in $\mathrm{BaTiO_{3}}$. In 
this treatment, there are five adjustable parameters to be fitted to the 
observed neutron intensity profile for each ${\bf q}$-value. It has been 
shown that, in order to reproduce the entire intensity distribution in 
the observed $q$-range, the effective width of TO-mode, $\Gamma_{1}(q)$, 
changes abruptly at ${\bf q}_{0}$  where `waterfall' takes place. They 
claim that the abrupt increase of $\Gamma_{1}$ would be caused by 
inhomogeneity of the lattice due to random distribution of PNR with 
characteristic size of $\sim q_{0}^{-1}$.

While this interpretation is very attractive, the basis of the `coupled mode' 
formalism on which the treatment is based, is a `homogeneous anharmonic 
lattice'. The damping of phonons are caused by the energy flow to the heat 
bath through anharmonic potential, whence the static heterogeneity of the 
medium would be out of the framework of the treatment.

We take somewhat different standpoint. Besides the phonon system, we 
introduce a stochastic variable whose dynamical behavior is only statistically 
determined. When such variable is strongly coupled to phonons, the life time 
of phonons would be mainly determined by the energy flow to the heat bath 
through the random variable. In perovskite relaxors, the configurational 
freedom of $\mathrm{Pb^{2+}}$ ions seems to play the role of such stochastic 
variable since it is well established\cite{Fujishiro} that the instantaneous 
equilibrium position of $\mathrm{Pb^{2+}}$ ion is slightly displaced from 
the corner site due to covalency effect, and it is randomly hopping across 
the potential barrier between the equivalent off-centered sites allowed by 
the averaged cubic symmetry m3m. We may describe this freedom of motion of 
$\mathrm{Pb^{2+}}$ ion by a pseudospin which takes on a few distinct values. 
Thus, our standpoint is schematically envisaged as illustrated in Fig.4.

\begin{figure}[h]
\begin{center}
\includegraphics[width=8.0cm,height=3.7cm]{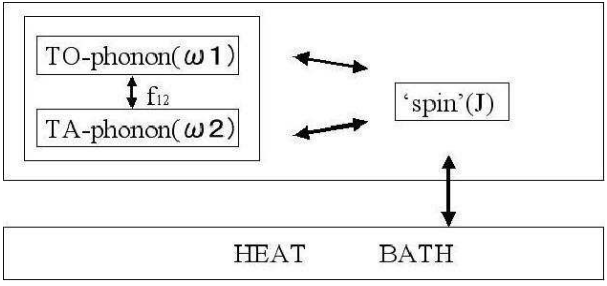}
\end{center}
\caption{Schematic description of the coupling scheme in 
TO-TA-pseudospin model. Major energy flow from the phonon system to heat 
bath is caused through the coupling to the pseudospin system.}
\label{fig:4}
\end{figure}

As the suitable framework to treat the coupled TO-TA-pseudospin system, we 
utilize Langevin equation which describes the motion of the variable under 
random force ${\bf f}(t)$,

\begin{eqnarray}
  \dot{\bf A}(t)={\bf \Omega}\cdot{\bf A}-\int{\bf \Phi}(t-t')\cdot
  {\bf A}(t')dt+{\bf f}(t),
\end{eqnarray}
                   
\begin{eqnarray}
  {\bf \Omega}=[\dot{\bf A},{\bf A}]\cdot{\bm \chi}^{-1},
\end{eqnarray}
                            
\begin{eqnarray}
  {\bf \Phi}=({\bf f}\cdot{\bf f}(t))\cdot{\bm \chi}^{-1},
\end{eqnarray}

\noindent where ${\bf A}$ is the state vector whose components are given 
by the independent variables of the system, ${\bm \chi}$ is the static 
susceptibility tensor defined by

\begin{eqnarray}
   \chi_{ij}=\int_{0}^{\beta}\langle A_{i}e^{-\lambda H}A_{j}e^{\lambda H}
   \rangle d\lambda.
\end{eqnarray}
                       
\noindent In the simple case of the random force with white spectrum, the time 
development of the averaged value of ${\bf A}(t)$ is given by\cite{Mori}
                         
\begin{eqnarray}
  \langle \dot{\bf A}(t)\rangle=({\bm \gamma}\cdot {\bm \beta})
  \langle A \rangle,
\end{eqnarray}
                         
with
                         
\begin{eqnarray}
  {\bm \gamma}=[\dot{\bf A},{\bf A}]+({\bf f}\cdot{\bf f}),
\end{eqnarray}
                             
\begin{eqnarray}
  {\bm \beta}={\bm \chi}^{-1}.
\end{eqnarray}
                             
\noindent Once the linear equation of motion is established as above it is not 
difficult to obtain the spectral representation of the correlation function, 
$\varphi_{ij}(\omega)=\int\langle A_{i}(t)A_{j}(0) \rangle e^{i \omega t}dt$, 
in the form\cite{Landau}: 

\begin{eqnarray}
  \varphi_{ij}(\omega)=[{\bm \zeta}+i \omega {\bm \beta}]_{ij}^{-1}
  +[{\bm \zeta}-i \omega {\bm \beta}]_{ij}^{-1},
\end{eqnarray}

\begin{eqnarray}
  {\bm \zeta}={\bm \beta}\cdot{\bm \gamma}\cdot{\bm \beta}.
\end{eqnarray}
                   
\noindent To apply the above general discussions to the present system, we 
define a five-component state vector:

\begin{eqnarray}
  {\bf A}^{+}=(P_{1}({\bf q}),Q_{1}({\bf q}),P_{2}({\bf q}),
  Q_{2}({\bf q}),  \sigma({\bf q})),
\end{eqnarray}

\noindent where $P_{1},Q_{1}$ are the momentum and the amplitude of the 
TO $(i=1)$ and TA $(i=2)$ phonons, and $\sigma({\bf q})$ is the Fourier 
transformed pseudospin variable:

\begin{eqnarray}
  \sigma({\bf q})=1/\sqrt{N}\int\sigma_{i}e^{i{\bf q}{\bf r}}dr.
\end{eqnarray}

\noindent As stated above, we assume that the random force, which is acting 
only on the pseudospin variable has the white spectrum: 

\begin{eqnarray}
  \langle ff(t) \rangle=\gamma \delta(t).
\end{eqnarray}
                           
\noindent Physically, $\gamma$ corresponds to the relaxation constant of the 
hopping motion of $\mathrm{Pb^{2+}}$-ion.

The energy of the coupled TO-TA-spin system is expressed by,

\begin{eqnarray}
 \lefteqn{H=\sum_{{\bf K}}\{ 
    \frac{1}{2}(P_{1}^{2}({\bf q})+
    \omega_{1}^{2}({\bf q})Q_{1}^{2}({\bf q}))}\nonumber\\  
     & +\frac{1}{2}(P_{2}^{2}({\bf q})+
     \omega_{2}^{2}({\bf q})Q_{2}^{2}({\bf q}))
     +\frac{1}{2}J\mid\sigma({\bf q})\mid^{2}\nonumber\\    
     & +f_{12}Q_{1}({\bf q})Q_{2}(-{\bf q})
     +g_{1}Q_{1}({\bf q})\sigma(-{\bf q})\nonumber\\
     & +g_{2}Q_{2}({\bf q})\sigma(-{\bf q}) 
  \},  
\end{eqnarray}
 
\noindent where the last three terms give the linear couplings between the 
spin and phonons.

Using eq. (16), we obtain the phonon spectral density in the 5-component 
system as follows:

\begin{eqnarray}
  S({\bf q},\omega)=\sum_{i=2,4}[{\bm \zeta}+i \omega {\bm \beta}]_{ii}^{-1}
  +[{\bm \zeta}-i \omega {\bm \beta}]_{ii}^{-1},  
\end{eqnarray}
                  
\begin{eqnarray}
  {\bm \zeta}={\bm \beta}\cdot{\bm \gamma}\cdot{\bm \beta},
\end{eqnarray}
                            
\noindent where ${\bm \beta}$ and ${\bm \gamma}$ are explicitly given 
(See Appendix) by, 
	
\begin{eqnarray}
    {\bm \beta}=\frac{1}{k_{B}T}
  \left(  \begin{array}{ccccc}
    1 & 0 & 0 & 0 & 0 \\
    0 & \omega_{1}^{2} & 0 & f_{12} & g_{1}\omega_{1} \\
    0 & 0 & 1 & 0 & 0 \\
    0 & f_{12} & 0 & \omega_{2}^{2} & g_{2}\omega_{2} \\
    0 & g_{1}\omega_{1} & 0 & g_{2}\omega_{2} & k_{B}T-J \\
  \end{array} \right),
\end{eqnarray}
	
\begin{eqnarray}
    {\bm \gamma}=
   \left(  \begin{array}{ccccc}
    0 & -1 & 0 & 0 & 0 \\
    1 & 0 & 0 & 0 & 0 \\
    0 & 0 & 0 & -1 & 0 \\
    0 & 0 & 1 & 0 & 0 \\
    0 & 0 & 0 & 0 & \gamma \\   
   \end{array} \right).
\end{eqnarray}
  
\noindent Notice if we eliminate the fifth row and column from ${\bm \beta}$ 
and ${\bm \gamma}$ tensors, the spectral density simply reproduces the `hypothetical' phonon dispersions given in Fig. 2 :

\begin{eqnarray}
S({\bf q},\omega)=\frac{1}{\hat{\omega}_{1}^{2}({\bf q})}
  \delta(\omega\pm\hat{\omega}({\bf q}))
  +\frac{1}{\hat{\omega}_{2}^{2}({\bf q})}
  \delta(\omega\pm\hat{\omega}_{2}({\bf q})),
\end{eqnarray}

\noindent where $\hat{\omega}({\bf q})$'s are the renormalized TO and TA phonon frequencies. In this context, the present treatment may be viewed as a 
natural extension of AHS formalism to include the pseudospin freedom 
of motion.

In order to visualize general characteristics produced by the coupling 
to pseudospin (stochastic) variable, we consider a simpler case of `single 
TO phonon-pseudospin coupled system.' In this case, we can obtain rather 
simple analytic expression of $S({\bf q},\omega)$ as follows

\begin{eqnarray}
  S({\bf q},\omega)=\frac{2k_{B}T\gamma\omega_{1}^{2}g^{2}}
  {\omega^{2}(\omega^{2}-\omega_{1}^{2})^{2}+\gamma^{2}\{g^{2}\omega_{1}^{2}
  +J'(\omega^{2}-\omega_{1}^{2})\}^{2}},
\end{eqnarray}
         
\begin{eqnarray}
  J'=k_{B}T-J
\end{eqnarray}
                           
\noindent Similar formula was already given by Yamada et 
al..\cite{Yamada2,Yamada3} It is worthwhile to notice that the profile of 
the TO phonon spectrum changes drastically as `two peak $\to $ broad single 
peak $\to $ triple peak' when the 
relative magnitude of $\omega_{0}$ and $\gamma$ is changed as 
$\gamma\gg \omega_{0}\to \gamma\sim\omega_{0}\to \gamma\ll \omega_{0}$. 
(See Figs. 1$\sim$3 in ref. 21). Hence, if the relaxation constant $\gamma$ 
satisfies the condition: $\gamma\sim\hat{\omega}_{TO}({\bf q}_{0})$, 
the profile of spectral distribution should exhibit the `waterfall'-like 
behavior around ${\bf q}\cong{\bf q}_{0}$.

Keeping these considerations in mind we use eq. (22) to calculate numerically 
the phonon spectral density distribution of PMN along [100]-direction in the 
region $0\leq q\leq 0.2$. The important parameter $\gamma$ has been taken 
as $\hbar\gamma=$7meV. The result is depicted in Fig. 5 (a), in comparison 
with the experimental results by Gehring et al.. In spite of that we have 
used only limited number of disposable parameters 
$(f_{12},g_{1}(=g_{2}),\gamma)$ the characteristic features of the observed 
neutron spectrum through the whole ${\bf q}$-region has been well reproduced.

\begin{figure*}[htb]
\begin{center}
\includegraphics[width=16.0cm,height=10.0cm]{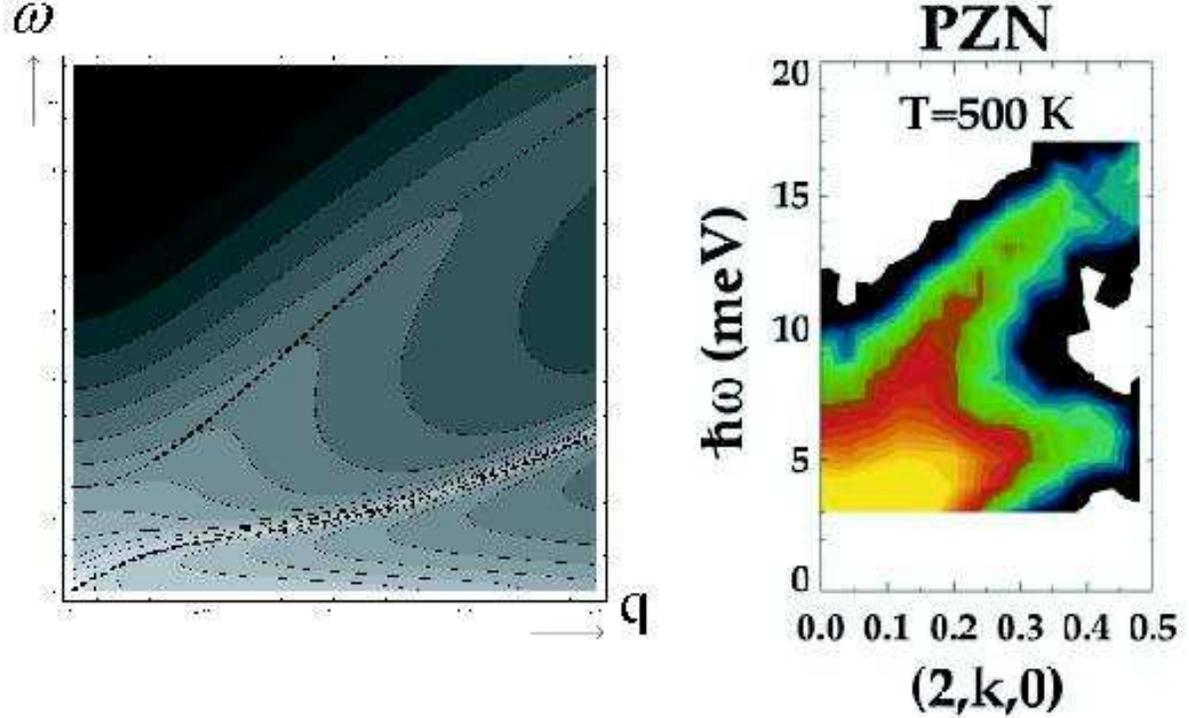}
\end{center}
\caption{(a) Calculated phonon spectral density distribution, 
$S({\bf q},\omega)$, along [100]-direction in PMN, which is compared with 
the observed neutron scattering spectrum by Gehring et al. \cite{Gehring2} 
reproduced in (b). The dashed lines corresponds to the `hypothetical' phonon 
dispersions when the coupling to the pseudospin is neglected.}
\label{fig:5}
\end{figure*}

\section{Conclusions and discussions}

In conclusion, we have developed a theory to study the characteristics of 
dipolar / strain fluctuations in perovskite relaxors. In addition to the soft 
TO and TA phonons, we take into account another freedom of motion associated 
with the random hopping of $\mathrm{Pb^{2+}}$ ions between the off-center 
sites around the high symmetry corner site, thus constructing a coupled 
TO-TA-pseudospin model to describe the perovskite relaxors.

It is shown that there is a possibility that prior to the on-set of 
instability of the uniform TO-mode (soft-mode), instability of TA mode 
with nanometer scale modulation $({\bf q}\sim 0.1{\bf a}^{*})$ takes place, 
which produces static heterogeneous structure concerning polarization as 
well as shear strain. In the case of PMN, the most probable direction of 
modulation is [110]-direction. This seems to suggest the intrinsic origin 
of the heterogeneity in relaxors visualized as random distribution of PNR.

The phonon spectral density distribution has also been investigated. It is 
shown that when the relaxation time of random hopping of $\mathrm{Pb^{2+}}$ 
ions between the equivalent off-center sites is comparable to the TO phonon 
frequency at ${\bf q}\cong 0.1{\bf a}^{*}$, the calculated phonon spectral 
density reproduces the characteristic features called `waterfall' which was 
observed by neutron spectra in PMN and PZN.\\
\\
\indent In this paper, we have discussed the possible origin of intrinsic 
heterogeneity in relaxors in terms of TA mode instability. Since the energy 
is simply assumed in a quadratic form, the resultant heterogeneity is 
expressed by a harmonic modulation. When higher order terms are taken 
into account, the modulation would become more or less `kink'-like 
producing well defined boundary between the polar and non-polar regions. 
Recently, Yamada and Takakura investigated the origin of IC phase and 
two-phase coexistence in $\mathrm{A_{0.5}B_{0.5}MnO_{3}}$ (CMR substance) 
observed around metal-insulator phase boundary based on TDGL formalism. 
Without the higher order coupling between 3d-orbital of $\mathrm{Mn^{3+}}$ 
ion and local strain, the system stabilizes a regular IC structure. However 
when the higher order electron-phonon coupling is taken into account, a 
two-phase coexistent state is stabilized in which metallic nanoregion is 
embedded randomly on the parent insulator phase.

Similar features would be expected in relaxors. Particularly, the random 
distribution of B-site ions provides the random pinning center to fix the 
`kink' (domain boundary) position, which will enhance the randomness of 
the spatial heterogeneity. This situation could explain the discrepancy 
between the observed and the calculated diffuse patterns around (h00)-type 
Bragg position as shown in Fig. 3.

\begin{acknowledgments}
     The authors would like to thank G. Shirane of Brookhaven National 
     Laboratory for valuable discussions and suggestions to draw our 
     attention to various important characteristics of relaxors.
\end{acknowledgments}


\appendix
\section{}
     We start with general thermodynamical equation of motion of a 
     mult-component variable ${\bf x}(=\{ x_{i} \})$:

\begin{eqnarray}
  \dot{\bf x}=-{\bm \gamma}{\bf X},
\end{eqnarray}

\noindent where ${\bf X}$ is the `driving force' of the system defined in terms of the thermodynamical potential $F({\bf x})$ given by

\begin{eqnarray}
  {\bf X}=\frac{1}{k_{B}T}\frac{\partial F}{\partial {\bf x}}.
\end{eqnarray}

\noindent When $F$ is given in the quadratic form with respect to ${\bf x}$ as, 
\begin{eqnarray}
  F=\sum_{il}\beta_{ik}x_{i}x_{l},
\end{eqnarray}

\noindent $X$  becomes the linear function of ${\bf x}$:

\begin{eqnarray}
  {\bf X}={\bm \beta}\cdot{\bf x}.
\end{eqnarray}

\noindent Substitution of (A4) into (A1) gives

\begin{eqnarray}
  \dot{\bf x}=-({\bm \gamma}\cdot{\bm \beta}){\bf x}.
\end{eqnarray}

\noindent Conversely, $\dot{\bf X}$ is expressed as

\begin{eqnarray}
  \dot{\bf X} & = & -{\bm \zeta}\cdot{\bf x} \nonumber \\
  & = & ({\bm \beta}\cdot{\bm \gamma}\cdot{\bm \beta}){\bf x}.
\end{eqnarray}

\noindent Using these linearized equations, the spectral representation 
of the two-time correlation function:

\begin{eqnarray}
  \varphi_{il}(\omega)=\int\langle x_{i}(t)x_{l}(t+\tau)\rangle 
  e^{i\omega\tau}d\tau
\end{eqnarray}

\noindent is expressed in terms of the coefficients ${\bm \beta}$, 
${\bm \gamma}$ as follows:

 \begin{eqnarray}
    \varphi_{il}(\omega)=[{\bm \zeta}-i \omega {\bm \beta}]_{il}^{-1}
   +[{\bm \zeta}+i \omega {\bm \beta}]_{li}^{-1}.
\end{eqnarray}                                         

\noindent We further notice that when $\varphi_{il}(\omega)$ is integrated 
over $\omega$, we obtain the instantaneous (same time) correlation as, 

\begin{eqnarray}
  \int\varphi_{il}(\omega)d\omega & = & \int\langle x_{i}(t)x_{l}(t+\tau)
  \rangle e^{i\omega\tau}d\tau d\omega \nonumber \\
  & = & \langle x_{i}x_{l}\rangle =k_{B}T{\bm \chi}_{il}
\end{eqnarray}

\noindent where ${\bm \chi}$ is the static susceptibility tensor. It is shown 
that within the linear approximation the instantaneous correlation is given by,

\begin{eqnarray}
  \langle x_{i}x_{l}\rangle =[{\bm \beta}^{-1}]_{il}.
\end{eqnarray} \\

In order to apply the above treatment to the present case, we consider that 
the long wave TA and TO phonons under consideration may be treated as 
thermodynamical variables. Using the energy given in eq. (21), the 
associated thermodynamical potential is expressed in a quadratic form;

\begin{eqnarray}
\lefteqn{F=\sum_{q}\frac{1}{2}(P_{1}^{2}({\bf q})
    +\omega_{1}^{2}Q_{1}^{2}({\bf q}))+\frac{1}{2}(P_{2}^{2}({\bf q})
    +\omega_{2}^{2}Q_{2}^{2}({\bf q}))} \nonumber \\
  & +\frac{1}{2}J_{k}\mid\sigma({\bf q})\mid^{2}-kT\mid\sigma({\bf q})\mid^{2}
    +f_{12}Q_{1}({\bf q})Q_{2}(-{\bf q}) \nonumber \\
  & +g_{1}\sigma({\bf q})Q_{1}(-{\bf q})
  +g_{2}\sigma({\bf q})Q_{2}(-{\bf q}).
\end{eqnarray}

\noindent As for the entropy contribution, we have only taken into account 
the configurational randomness of the pseudospin variables.

Finally we compare the fundamental equations (13), (14), and (15) in the 
text with (A5), (A9) and (A10), from which the expressions for 
${\bm \beta}$-and ${\bm \gamma}$-tensors are deduced as given in eqs. (24) 
and (25).

\newpage

%



\begin{thebibliography}{99}
\bibitem{dissipation}	In the present treatment the parameter to describe 
the energy dissipation is solely included in the ${\bm \gamma}$-matrix defined 
in eq. (25). Since the instantaneous correlation is given in terms of 
${\bm \beta}$-matrix (See eq. (A10) in Appendix.), the X-ray intensity is 
irrelevant to the damping mechanism.
\end{thebibliography}
\end{document}